\documentclass[accepted]{melba}

\usepackage{amsmath,amsfonts}
\usepackage{tabularx}
\usepackage[final]{changes}

\melbaid{2026:029}  
\doi{10.59275/j.melba.2026-d29c}
\melbaauthors{Heyer et al}  
\email{w.heyer@uni-luebeck.de}
\volume{2026}
\firstpageno{594}  
\melbayear{2026}  
\datesubmitted{2025-07-21}  
\datepublished{2026-10-11}  

\ShortHeadings{OncoReg}{Heyer et al.}

\title{OncoReg: Medical Image Registration for Oncological Challenges}

\author{
    \firstname Wiebke \surname Heyer\aff{1}\orcid{0009-0007-1147-8770},
    \firstname Yannic \surname Elser\aff{2}\orcid{0009-0004-1267-4202}, 
    \firstname Lennart \surname Berkel\aff{2}\orcid{0009-0006-8024-6992},
    \firstname Xinrui \surname Song\aff{3}\orcid{0000-0001-5570-1056}, 
    \firstname Xuanang \surname Xu\aff{3}\orcid{0000-0002-6045-8457},
    \firstname Pingkun \surname Yan\aff{3}\orcid{0000-0002-9779-2141},
    \firstname Xi \surname Jia\aff{4}\orcid{0009-0004-9815-3706},
    \firstname Jinming \surname Duan\aff{4,5}\orcid{0000-0002-5108-2128},
    \firstname Zi \surname Li\aff{6}\orcid{0000-0001-7952-7021},
    \firstname Tony C. W. \surname Mok\aff{6}\orcid{0000-0002-4779-9337},
    \firstname Bowen \surname Li\aff{7},
    \firstname Tim \surname Hable\aff{1},
    \firstname Christian \surname Staackmann\aff{8},
    \firstname Christoph \surname Großbröhmer\aff{1}\orcid{0000-0002-8926-8729},
    \firstname Lasse \surname Hansen\aff{9}\orcid{0000-0003-3963-7052},
    \firstname Alessa \surname Hering\aff{10}\orcid{0000-0002-7602-803X},
    \firstname Malte M. \surname Sieren\aff{2,11}\orcid{0000-0002-3669-2892},
    \firstname Mattias P. \surname Heinrich\aff{1}\orcid{0000-0002-7489-1972}
}

\affiliations{
	\num 1 \addr Institute of Medical Informatics, University of Lübeck, Lübeck, Germany\\
	\num 2 \addr Institute of Radiology and Nuclear Medicine, University Hospital Schleswig-Holstein, Lübeck, Germany \\
	\num 3 \addr Department of Biomedical Engineering and Center for Biotechnology and Interdisciplinary Studies, Rensselaer Polytechnic Institute, Troy, NY, USA\\
    \num 4 \addr School of Computer Science, University of Birmingham, Birmingham, UK\\
    \num 5 \addr Division of Informatics, Imaging and Data Sciences, University of Manchester, UK\\
    \num 6 \addr DAMO Academy, Alibaba Group, Hangzhou, China\\
    \num 7  \addr Hangzhou Shengshi Technology Co., Ltd, Hangzhou, China\\
    \num 8 \addr Department of Radiation Oncology, University Hospital Schleswig-Holstein, Lübeck, Germany\\
    \num 9 \addr EchoScout GmbH, Lübeck, Germany\\
    \num 10 \addr Radboud University Medical Center, Nijmegen, Netherlands\\
    \num 11 \addr Institute of Interventional Radiology, University Hospital Schleswig-Holstein, Lübeck, Germany\\
}

\abstract{
    In modern cancer research, the vast volume of medical data generated is often underutilised due to challenges related to patient privacy. The OncoReg Challenge addresses this issue by enabling researchers to develop and validate image registration methods through a two-phase framework that ensures patient privacy while fostering the development of more generalisable AI models. Phase one involves working with a publicly available dataset, while phase two focuses on training models on a private dataset within secure hospital networks. OncoReg builds upon the foundation established by the Learn2Reg Challenge by incorporating the registration of interventional cone-beam computed tomography \deleted{(CBCT)} with standard planning fan-beam CT \deleted{(FBCT)}images in radiotherapy. Accurate image registration is crucial in oncology, particularly for dynamic treatment adjustments in image-guided radiotherapy, where precise alignment is necessary to minimise radiation exposure to healthy tissues while effectively targeting tumours.\\
    This work details the methodology and data behind the OncoReg Challenge and provides a comprehensive analysis of the competition entries and results. Findings reveal that feature extraction plays a pivotal role in this registration task. A new method emerging from this challenge demonstrated its versatility, while established approaches continue to perform comparably to newer techniques. Both deep learning and classical approaches still play significant roles in image registration, with the combination of methods, particularly in feature extraction, proving most effective.
    }

\keywords{Image Registration, Cone-beam Computed Tomography, Image-Guided Radiotherapy, Computational Challenge}

\begin{document}

\twocolumn[\maketitle]

\section{Introduction}
    \enluminure{I}{n radiology}, and particularly in modern cancer research, vast amounts of data are continuously generated. However, much of this data remains underutilised for developing innovative AI-centred solutions due to several challenges. One significant obstacle is the restriction on data publication imposed by patient privacy rights. This is especially problematic for data collected in previous years, where obtaining retroactive patient consent is no longer feasible. To unlock the potential of such data while ensuring compliance with privacy laws, novel approaches are required. 
    The OncoReg Challenge addresses this issue by enabling researchers to develop and validate methods in a two-phase framework. In the first phase, challenge participants work with a publicly available dataset to design and refine their algorithms. In the second phase, the submitted models are trained in-house on a private, unpublished dataset\added{,} ensuring that the data remains securely within the hospital network and never becomes accessible to external participants.
    
    This approach not only safeguards patient privacy but also fosters the development of more generalisable AI models. Since participants must optimise their methods for unseen data rather than a specific dataset, the challenge encourages the creation of algorithms with broader applicability and better real-world transferability. Computational challenges provide an ideal framework for this kind of data utilisation, allowing multiple research groups to benefit from the dataset without the need for public distribution. The popularity of challenges, especially in the field of computer vision, has been steadily increasing, as reflected in the growing number of challenges hosted at major conferences such as MICCAI, ISBI, and SPIE Medical Imaging. 

    \added{The Learn2Reg Challenge (\citealp{hering2022learn2reg}) has established itself as a leading benchmark in the field of medical image registration.} The well-known challenge, which has been held annually since 2019 in conjunction with MICCAI, has significantly advanced algorithmic approaches and benchmarking in medical image registration, fostering collaboration and innovation among researchers worldwide. Learn2Reg offers an extensive selection of 3D image registration tasks, covering a variety of scenarios. The tasks include both mono- and multimodal registration as well as inter- and intrapatient registration. The challenges span multiple imaging modalities such as computed tomography (CT), magnetic resonance imaging (MRI), and ultrasound (US), and encompass different anatomical regions, including the lungs, abdomen, and brain. 
    OncoReg builds upon the foundation established by Learn2Reg, extending it to include an additional modality, cone-beam CT, and shifting the focus to real-world oncological challenges. The first part of OncoReg comprises a task (ThoraxCBCT) in the Learn2Reg Challenge 2023, followed by a standalone virtual challenge and workshop.
    
    This focus on real-world scenarios highlights the importance of image registration, which has numerous medical applications but plays an especially critical role in oncology, particularly in the context of radiotherapy, where precise targeting of cancerous tissues is vital for effective treatment. Radiotherapy involves delivering high doses of radiation to tumour sites while minimising exposure to surrounding healthy tissues. To achieve this level of precision, accurate alignment of images from different time points and imaging modalities is crucial.
    
    In particular, the registration of low-dose interventional cone-beam CTs (CBCTs) with high-resolution diagnostic fan-beam CTs (FBCTs) is a fundamental step in image-guided radiotherapy (\citealp{zhang2024bcswinreg, cao2022cdfregnet, han2021deep}). FBCTs, acquired before treatment, provide a detailed map of the patient’s anatomy and are used to plan the radiation dose distribution. However, during the course of treatment, the patient’s anatomy can change due to tumour shrinkage, weight loss, or other factors. CBCTs are frequently acquired during treatment sessions to assess these changes in real time. By registering CBCTs with the initial planning CTs, clinicians can adjust the treatment plan dynamically, ensuring that radiation continues to be delivered precisely to the target area, even as the anatomy evolves.
    
    This registration process is technically challenging due to differences in image quality, patient positioning, and anatomical changes over time. Therefore, developing robust algorithms for accurate image registration is essential to improving the efficacy of radiotherapy. Standardising planning processes and continuously enhancing registration techniques, particularly with the support of adaptive algorithms, not only enhances patient safety but also minimises treatment-related side effects. The development and refinement of such algorithms through challenges like OncoReg contribute to optimising treatment options for cancer patients.
    
    This work presents the methodological approach used in conducting the OncoReg challenge and provides a detailed description and analysis of the entries, results, and rankings.

\section{Related Work}

    \subsection{Image Registration Challenges}
    The EMPIRE10 challenge (\citealp{murphy2011evaluation}) was one of the first widely recognised image registration challenges, specifically targeting the elastic nature of lung tissue deformations. The challenge evaluated registration performance on 30 thoracic CT scan pairs using several metrics: distances between manual landmark pairs, fissure segmentations, and Jacobian determinant values of the deformation field. A unique feature of the original 2010 workshop was the requirement for live computations, placing time constraints on the registration algorithms. EMPIRE10 set a foundational precedent for evaluating pulmonary image registration and established performance metrics that have influenced subsequent challenges.
    
    Several years later, the Continuous Registration Challenge (\citealp{marstal2019continuous}) introduced a continuous benchmarking system with automatic evaluation based on eight datasets, focusing on lung CT and brain MRI registration. However, the framework was somewhat limited by the requirement that algorithms had to be compatible with the SuperElastix platform. In 2019, two other major challenges emerged alongside the first iteration of the Learn2Reg Challenge (\citealp{hering2022learn2reg}). The CuRIOUS challenge (\citealp{xiao2019evaluation}) addressed brain shift correction using intraoperative ultrasound and preoperative MRI, while the ANHIR challenge (\citealp{borovec2020anhir}) focused on registering histological images. In the following years, BraTS-Reg (\citealp{baheti2021brain}) provided a benchmark for registering preoperative and follow-up MRI scans in diffuse glioma patients, and the FetReg challenge (\citealp{bano2021fetreg}) centred on placental vessel registration. More recently, the ACROBAT challenge (\citealp{weitz2024acrobat}) aimed to improve the alignment of stained whole-slide images from breast cancer tissue sections.
    
    While many registration challenges have focused on specific tasks or imaging modalities, Learn2Reg stands out for its comprehensive approach. It covers a wide range of tasks, including mono- and multimodal registration, across diverse anatomical regions and imaging modalities such as CT, MRI, and ultrasound. With the addition of OncoReg to Learn2Reg in 2023 in the form of the ThoraxCBCT task, the challenge introduced its first oncological lung registration task. This also marked the inclusion of cone-beam computed tomography \deleted{CBCT} alongside conventional diagnostic fan-beam CT\deleted{FBCT}, expanding the scope of the challenge to address real-world clinical problems in oncology.

    \subsection{Cone-Beam CT Registration}
    
    The registration of CBCTs has various clinical applications\deleted{in medicine}, with image-guided radiotherapy being one of the most prominent (\citealp{zhang2024bcswinreg, cao2022cdfregnet, han2021deep}). A related application is image-guided surgery, \added{with video-assisted thoracoscopic surgery (VATS) representing a more specialised application, a minimally invasive technique for lung nodule resection in early-stage lung cancer (\citealp{alvarez2021hybrid})}. Beyond oncology, CBCT registration is also utilised in dental imaging (\citealp{zheng2025automatic}) and robotic cochlear implant surgery (\citealp{lu2022preoperative}).
    
    Despite the widespread use of CBCT, its registration to conventional CT remains a challenging problem due to lower image quality, artefacts, and intensity variations. While general-purpose registration methods have been evaluated on CBCT-CT tasks (\citealp{demir2024multigradicon, van2024reindir}), dedicated solutions remain relatively scarce. Historically, semi-automatic optimisation-based methods have been applied (\citealp{sotiras2013deformable}), and classical optimisation approaches continue to be refined (\citealp{xie2022multi}). More recently, deep learning-based methods have emerged, such as an unsupervised dual-attention network (\citealp{hu2022unsupervised}) and a rigid deep-learning-based registration method for head-neck CT-CBCT (\citealp{peng2024acswinnet}). Additionally, \cite{teng2021respiratory} introduced a supervised CNN for phase-to-phase registration in 4D CBCT, and \cite{li2023incorporating} proposed generating high-quality synthetic CTs from CBCTs as an alternative approach to facilitate registration.
    
    While progress has been made, CBCT-CT registration remains an open research challenge, with ongoing efforts to develop more robust, artefact-resistant, and anatomically aware methods that can improve clinical workflows and treatment outcomes.
    
\section{Challenge Organisation}
        
    The OncoReg challenge was split into two phases, the first of which was organised as part of the 2023 version of the Learn2Reg challenge associated with MICCAI 2023 and the second as MICCAI-endorsed virtual workshop in February 2024. The Learn2Reg challenge was carried out as a Type 1 and Type 2 challenge whereas the second phase resulting in the OncoReg workshop was carried out as a Type 3 challenge. The OncoReg challenge was organised by Mattias Heinrich, Alessa Hering, Christoph Großbröhmer and Wiebke Heyer. 
    
    Computational challenges in the field of machine learning are often categorised based on the accessibility of data to participants. These challenges generally fall into four distinct types.
    Type 0 challenges offer participants access to all data phases, including supervision data, which minimises the workload for organisers. However, this approach compromises fairness, as participants are responsible for computing their own metrics, leading to potential inconsistencies.
    Type 1 challenges improve on this by keeping the supervision data for the test phase private, while the rest of the data is publicly accessible. Participants submit their predictions, which the organisers then evaluate and rank. This structure ensures a more consistent evaluation process but still allows participants access to much of the data.
    Type 2 challenges further increase the complexity by withholding not just the supervision data but also the test phase data itself. Participants must submit an inference algorithm, which the organisers run on the hidden data before evaluating and ranking the results. This approach increases the workload for both organisers and participants. Participants often need to containerise their methods or compile them into specific formats, adding to the complexity.
    Type 3 challenges represent the most complex scenario, where both the training and test data are hidden. Participants must submit their entire training algorithm, which organisers then use to retrain models on the hidden data before conducting inference, evaluation, and ranking. While this increases the burden on organisers significantly, it offers several advantages. One key benefit is that sensitive data, such as medical data that cannot be publicly disclosed due to patient privacy concerns, can still be utilised for research and algorithm development. Additionally, this approach ensures that all submissions are evaluated under identical conditions, eliminating the use of additional data, annotations, or varied hardware, thereby drastically enhancing fairness.
    
    Furthermore, keeping data private in this manner \added{can encourage the development of more generalisable solutions by reducing opportunities for manual adaptation to a specific dataset. However, the extent to which a challenge evaluates generalisation depends on the similarity between the hidden training and test domains and the availability of auxiliary data from related domains.}
    
    \added{It should be noted that challenge type definitions describe only data accessibility and evaluation procedures. They are independent of the relationship between training and evaluation domains. Depending on the challenge design, hidden test data may originate from the same distribution as the training data, from a different institution, scanner, acquisition protocol, or even a substantially different domain. Consequently, the ability of a challenge to assess model generalisation depends not only on the challenge type itself but also on the degree of domain shift between training and evaluation data.}
    
    \added{In the Learn2Reg Type 1 and Type 2 challenge, training and test cases originated from the same dataset and therefore shared the same imaging distribution, institution, and acquisition characteristics. In contrast, a domain shift existed between the public Learn2Reg dataset and the hidden OncoReg dataset used for the Type 3 workshop, as the latter was acquired at a different institution using different scanners and clinical acquisition protocols. Within the OncoReg dataset itself, however, no intentional domain shift was introduced between training and test cases. Participants therefore had access to the publicly available Learn2Reg data as a proxy source domain, while training and evaluation were performed on a hidden clinical target domain of the same imaging modality but different origin.}

    \subsection{Learn2Reg 2023 ThoraxCBCT Task}
        In 2023, the Learn2Reg challenge expanded its scope by introducing the new ThoraxCBCT task and extending the National Lung Screening Trial (NLST) task from the 2022 iteration. It also continued previous tasks such as LungCT, OASIS, and AbdomenMRCT, building on the challenge’s comprehensive history (\citealp{hering2022learn2reg}). The two primary tasks of the competition were hosted entirely on grand-challenge.org and followed a structured, two-phase process:
        \begin{itemize}
            \item Validation Phase: During this phase, participants developed their algorithms using the provided training and validation data. For evaluation, they were required to compute displacement fields for the validation cases and upload them to grand-challenge.org. Participants were allowed multiple submissions per day, with results and rankings updated in real-time on a public leaderboard. This encouraged iterative improvement and competition among the teams.
            \item Test Phase: Participants could either request access to the test data, submitting their computed displacement fields as in a traditional Type 1 challenge, or they could opt to submit their containerised algorithms through the grand-challenge.org/algorithms platform, marking a Type 2 challenge. This second option, which promoted algorithm submissions, was highly encouraged by the organisers to support the publication of user-friendly, ready-to-use methods on the platform. An additional evaluation of the inference runtime was conducted for those submitting containerised algorithms, emphasising the importance of both accuracy and efficiency in real-world applications.
        \end{itemize}
    \subsection{OncoReg Workshop}
        The second phase of OncoReg expanded on the Learn2Reg challenge, building upon the ThoraxCBCT task and transforming it into a Type 3 challenge that utilised a completely hidden, unpublished dataset. The OncoReg workshop placed a special emphasis on the role of registration within the context of radiotherapy.
        
        Participants were encouraged to develop new methods or adapt their Learn2Reg submissions for the Type 3 task, using the ThoraxCBCT leaderboard for initial validation. They were then given the opportunity to submit their containerised solutions for a preliminary sanity check. For deep learning-based approaches, final submissions were retrained and evaluated on the hidden dataset. Due to privacy restrictions on the dataset, this phase of the challenge was not conducted on grand-challenge.org.

    \subsection{Lowering Entry Barriers}
        To keep the challenge accessible to potential participants with limited experience in image registration and containerisation, \added{several measures were implemented to lower the entry barrier.} The different steps of participation were accompanied in several ways: To help participants in getting started in their algorithm development pre-processed data was provided and baseline methods for comparison were made public on the leaderboards. Code for evaluation including multiple metrics was published and examples for the creation of an algorithm on grand-challenge.org were provided on GitHub and grand-challenge.org. Regarding the Type 3 challenge an example submission docker with step-by-step instructions was provided and sanity checks were offered. 
        Additionally, participants had the opportunity to ask questions and discuss in a public Q\&A session, via a discussion forum and via direct email to the organisers. All additional resources for Learn2Reg 2023 and OncoReg can be found on our grand-challenge website\footnote{\href{https://learn2reg.grand-challenge.org/oncoreg/}{learn2reg.grand-challenge.org/oncoreg/}} and our GitHub\footnote{\href{https://github.com/MDL-UzL/L2R/}{github.com/MDL-UzL/L2R/}, \href{https://github.com/MDL-UzL/OncoReg/}{github.com/MDL-UzL/OncoReg/}}.

\section{Tasks \& Data}

        \begin{table*}[ht]
        \centering
        \caption{Overview of the two datasets \deleted{used} for the ThoraxCBCT and OncoReg tasks.}
        \begin{tabularx}{\linewidth}{l|X|X}
        \textbf{Dataset}  & \textbf{ThoraxCBCT}           & \textbf{OncoReg}    \\ \hline
        Modality & Fan-beam CT \& Cone-beam CT & Fan-beam CT \& Cone-beam CT \\
        4D (max. Insp./Exp.) & \checkmark & $\times$ \\
        \# Cases (2 Pairs/Case) & 19 (1xFBCT, 2xCBCT per Case) & 27 (1xFBCT, 2xCBCT per Case) \\
        Split & 11 Train, 3 Val., 5 Test & 3-Fold CV (15 Train, 3 Val., 9 Test) \\
        Image Size & 390, 280, 300 & 256, 192, 192 \\
        Spacing & 1~mm isotropic & 1.5~mm isotropic \\
        Manual Pre-alignment & \checkmark & \checkmark \\
        Label Masks & Semi-automatic, 13 Structures & Semi-automatic, 52 Structures \\
        Landmark Annotations & Manual, Avg. 16 per Image & Manual, Avg. 25 per Image \\
        Binary Trunk Masks & \checkmark & \checkmark \\
        Paired Förstner Keypoints & \checkmark (DeedsBCV) & \checkmark (ConvexAdam)\\
        Access & Original Data publicly available via TCIA, Processed Data available on Challenge Website & Private Dataset, Evaluation on data per request \\
        
        \end{tabularx}
        \label{tab:data}
        \end{table*}

        \subsection{ThoraxCBCT}
            With the release of the ThoraxCBCT dataset the registration problem of image-guided radiotherapy between pre-therapeutic \deleted{fan-beam CT (}FBCT\deleted{)} and interventional low-dose \deleted{cone-beam CT (}CBCT\deleted{)} is addressed. The released image data is part of the 4D-Lung dataset (\citealp{hugo2016data}) from The Cancer Imaging Archive which contains four-dimensional lung images acquired during radiochemotherapy of locally-advanced, non-small cell lung cancer (NSCLC) patients.
            
            For each patient one FBCT prior to therapy in maximum inspiration and two CBCTs, one at the beginning of therapy and one at the end of therapy, both in maximum expiration, are provided. The task is to find a solution for the registration of two pairs of images per patient:
            \begin{enumerate}
                \item The planning FBCT (inspiration) prior to therapy and the low dose CBCT at the beginning of therapy (expiration), which are acquired at similar timepoints, and
                \item The planning FBCT (inspiration) prior to therapy and the low dose CBCT at the end of therapy (expiration), where longer periods of time (usually several months) \added{elapsed} between the scans.
            \end{enumerate}
            The registration challenge in both image pairs involves aligning two different modalities: Fan-Beam CT and Cone-Beam CT, while accounting for shifts between maximum inspiration and expiration phases of breathing. The second subtask presents an additional challenge with temporal differences, as it compares the planning CT taken at the start of therapy with the follow-up CBCT acquired at the end of therapy.
            
            The dataset released includes training images from 11 patients, comprising 11 FBCT and 22 CBCT scans, and validation images from 3 patients, which consist of 3 FBCT and 6 CBCT scans. Each patient study includes two image pairs, resulting in a total of 22 training pairs and 6 validation pairs. For every image pair, Förstner keypoints are provided, based on displacement fields computed using \textit{DeedsBCV} (\citealp{deeds1}). Unlike previous Learn2Reg tasks that focused primarily on the lungs, this task extends to include other thoracic structures, particularly organs at risk in radiotherapy. Keypoints were therefore determined not only within the lungs but across the entire trunk region.
            
\begin{figure}[h]
            		  \centering
            		  \includegraphics[width=1\linewidth]{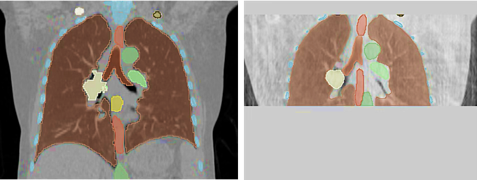}
            		  \caption{Exemplary image pair from the ThoraxCBCT dataset including semi-automatic label mask annotations. Left: Fan-Beam CT. Right: Cone-Beam CT.}
                      \label{fig:ThoraxCBCT_example}
            \end{figure}

            The test dataset includes images from 5 patients, yielding 5 FBCT and 10 CBCT scans (10 test pairs). The validation and test annotations feature multiple segmentation masks for various structures, including lungs, tumours, organs at risk (heart, spinal cord, oesophagus, trachea, and aorta), and skeletal structures such as vertebrae, ribs, clavicles, and scapulae. Additionally, anatomical landmarks like airway bifurcations, the top of the aortic arch, apex cordis, and key points in bone structures were marked. Some annotations for FBCTs were sourced from TCIA, with missing segmentation masks generated using the TotalSegmentator tool (\citealp{wasserthal2023totalsegmentator}) and propagated to CBCTs through registration. These generated masks were validated by clinicians at the University Hospital of Schleswig-Holstein (UKSH). On average, 13 structures were segmented per image. Landmark annotations were entirely manual, resulting in an average of 16 landmarks per image pair. A binary body-trunk mask, distinguishing the thorax from the background, was provided for each image, also created using the TotalSegmentator. An example of an image pair of the ThoraxCBCT dataset including label mask annotations is given in Fig.~\ref{fig:ThoraxCBCT_example}.
            
            All images were converted from DICOM to NIfTI format and subsequently resampled and cropped to the thoracic region, resulting in an image size of 390x280x300 with a voxel spacing of 1.0x1.0x1.0\,mm. Large translations between image pairs were manually corrected prior to registration.
            
            Training and validation images, trunk masks and keypoints are available via our grand-challenge website and test images as well as all supervision data is available upon request or via automatic evaluation on grand-challenge.org. 

        \subsection{OncoReg}
            The OncoReg task extends the ThoraxCBCT challenge but is conducted as a Type 3 task using a completely hidden dataset. Participants are tasked with registering a high-resolution diagnostic FBCT acquired before therapy with two CBCT scans: one from the start and another from the end of therapy. Unlike the ThoraxCBCT data, OncoReg images are not synchronised with respect to breathing phases, meaning the scans may capture different states of inspiration and expiration. The data, collected from routine clinical care at the University Hospital of Schleswig-Holstein (UKSH), was retrospectively curated from 27 lung cancer patients and carefully annotated with semi-automatic segmentations and manual landmarks.
            Similar to ThoraxCBCT, trunk masks and Förstner keypoints were computed for each pair. Preprocessing included converting images from DICOM to NIfTI format, cropping, resampling to a size of 256x192x192, with a voxel spacing of 1.5x1.5x1.5\,mm, and manual pre-alignment of image pairs. 

\begin{figure}[h]
            		  \centering
            		  \includegraphics[width=1\linewidth]{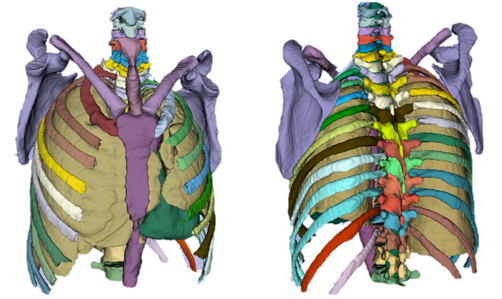}
            		  \caption{Semi-automatic label mask annotations for an exemplary Fan-Beam CT from the OncoReg dataset. 3D rendering generated with 3D Slicer (\citealp{fedorov20123d}).}
                      \label{fig:OncoReg_example}
            \end{figure}

            Annotations for OncoReg were more comprehensive than those for ThoraxCBCT, as the entire dataset was annotated, rather than just a subset. Segmentations were created using the TotalSegmentator tool and then manually refined, resulting in an average of 52 annotated structures per image. Manual landmark annotation was also more detailed, with an average of 25 landmarks identified per image pair. An overview of the label mask annotations is given in Fig.~\ref{fig:OncoReg_example}.
                    
            While this dataset is not publicly available, researchers may request evaluation of their registration methods on the data via the challenge platform.
            An overview of the two datasets is given in Table~\ref{tab:data}.

        \subsection{Challenge Design}
            \added{The ThoraxCBCT train, validation, and test split was not generated randomly. Instead, cases were selected to ensure that the validation and test sets reflected the variability present in the dataset with respect to image quality, noise level, deformation magnitude, patient anatomy (regarding gender and weight), and CBCT field-of-view. At least one particularly challenging case was included in the test set. While the resulting number of test cases is relatively small, all publicly available 4D-Lung cases with valid FBCT-CBCT pairings and sufficient image quality were utilised. A larger-scale cross-validation was not feasible within the Learn2Reg challenge framework, which relies on a fixed benchmark split and partially hidden test data for fair comparison between participating methods.}
        
            To ensure a comprehensive evaluation of the registration performance, a range of complementary metrics are employed that are adapted after previous editions of Learn2Reg and are outlined in Table~\ref{tab:metrics}. These metrics \added{were} chosen to evaluate various aspects of algorithmic performance, including accuracy, robustness, plausibility, and computational efficiency. The final ranking also accounts for the statistical significance of the differences between results, using a ranking system inspired by the Medical Decathlon (\citealp{antonelli2022medical}). Methods are statistically compared using the Wilcoxon signed-rank test, with a significance level set at $p<0.05$. Based on the number of won comparisons, each method receives a numerical rank score ranging from 0.1 to 1. The overall task ranking is then determined by calculating the geometric mean of these individual metric scores.
    
            \begin{table*}[ht]
            \centering
            \caption{Summary of Metrics used throughout the challenge for evaluation of registration performance.}
            \begin{tabularx}{0.75\linewidth}{c|X|X|c}
     
            \textbf{Metric}  & \textbf{Description}           & \textbf{Purpose}                                & \textbf{Unit}  \\ \hline
            DSC              & Dice Similarity Coefficient    & Overlap between segmentations                   & \%             \\ 
            DSC30            & 30th Percentile of DSC         & Measures robustness of overlap                  & \%             \\ 
            HD95             & 95th Percentile Hausdorff Dist. & Robust surface distance                       & mm            \\ 
            TRE              & Target Registration Error      & Landmark misalignment                           & mm           \\ 
            TRE30            & 70th Percentile of TRE         & Robustness of landmark error                    & mm        \\ 
            SDlogJ           & Standard Deviation of log(Jac.) & Local volume changes                             & -              \\ 
            RT               & Runtime                        & Time taken for registration                     & s               \\
            \end{tabularx}
            \label{tab:metrics}
            \end{table*}

\section{Challenge Entries}
    The ThoraxCBCT validation leaderboard featured 68 entries from seven teams (including the challenge organisers), with seven valid submissions made during the test phase. Four of these submissions were baseline methods provided by the organisers.
    The Type 3 OncoReg task received five submissions from four participating teams, in addition to five baseline methods supplied by the organisers. The baseline methods included \textit{NiftyReg} (\citealp{niftyreg09}), \textit{DeedsBCV} (\citealp{deeds1, Heinrich2015MultimodalMS}), \textit{ConvexAdam} (\citealp{siebert2021fast, heinrichconvex14}), and \textit{VoxelMorph++} (\citealp{heinrich2022voxelmorph++}). \textit{VoxelMorph++}, in particular, was made available as an algorithm on grand-challenge.org/algorithms and was containerised as an example on the challenge’s GitHub repository, which is why it is discussed in detail in the following section. Additionally, a keypoint based registration method using the medical open network for AI (MONAI) (\citealp{cardoso2022monaiopensourceframeworkdeep}) is provided. The baseline methods will not be further elaborated upon here, as they are already well-documented in existing literature. The methods submitted by the participating teams are described in detail below.
    
    \subsection{DINO-Reg \& DINO-Reg Ensemble} 
    DINO-Reg (\citealp{dinoreg_miccai}) is an optimisation-based method for medical image registration. The process begins with the encoding of medical image features using DINOv2 (\citealp{oquab2024dinov}), followed by an optimisation step applied to these encoded features to achieve accurate registration. DINOv2 is a 2D vision transformer pre-trained on natural images without labels, using knowledge distillation. The model demonstrates a strong understanding of image semantics and exceptional generalisability, making it suitable for medical image feature encoding without the need for fine-tuning. To adapt the model for 3D medical image feature extraction, encoding is performed on axial view slices, and the most relevant principal components are extracted from all features within the 3D volume. The resulting features facilitate accurate registration via simple gradient descent optimisation.
    
    Specifically, each 2D slice from the image is encoded into a 3D feature map using DINOv2, with the third dimension representing the feature space. Stacking the 3D feature maps from all slices generates a 4D feature map of the entire volume. Once the 4D feature maps for both the moving and reference images are obtained, principal component analysis (PCA) is performed on all feature tokens to (1) reduce feature dimensionality, and (2) align the moving and reference image features into the same feature space. The principal components aim to capture the variance within the original features. For these features to be meaningful in the context of medical image registration, the principal components must describe anatomical differences, such as those between organs and bones. If left unprocessed, most of the variance would arise between foreground and background content. Therefore, thresholding is applied to the original image intensity to isolate foreground patch features.
    
    Due to the patch-encoding step of vision transformers, the resulting feature map has a resolution 14 times lower than the original image. To mitigate this effect, the original image is up-sampled prior to encoding. During implementation, feature interpolation between slices is applied, and PCA is replaced with low-rank PCA to accelerate the method. For the CBCT data, DINO-Reg features and MIND (modality-independent neighbourhood descriptor; \cite{Heinrich12MIND}) features are found to be complementary, with the former providing semantic information and the latter capturing edge details. Consequently, a second round of optimisation using MIND features was tested after performing DINO-Reg, resulting in a method named DINO-Reg Ensemble.

    \subsection{Fourier-Net} 
        U-Net-based methods have dominated medical image registration due to their fast inference speed. The deep models take the full-resolution images (\(I_M, I_F\)) as input and estimate full-resolution displacements (\(\Phi\)). To reduce the repeated convolution operations in the U-Net architecture and boost computational efficiency, a model-driven method called Fourier-Net is proposed to learn a low-dimensional representation of the displacement.
        
        Fourier-Net operates under the hypothesis that the desired displacement is smooth, allowing it to be represented using only the low-frequency components. However, it has been experimentally demonstrated in \cite{Jia23} that learning the complex-valued frequencies of displacement from spatial images is challenging for a convolutional network. Fourier-Net instead learns a small spatial patch (\(\mathbb{S}_\phi\)) that shares the same low frequencies with the desired full displacement. Theoretically, once the small patch is perfectly learned by the network, incorporating the fast Fourier transform (FFT) transform, zero-padding, and inverse FFT transform will get the final displacement:
        \begin{align}
            \Phi = iFFT(Pad(FFT(\mathbb{S}_\phi)))
            \label{eq:fouriernet}
		\end{align}
        However, Fourier-Net only learns the low-frequency components of the displacement, therefore\added{, the estimated} deformation fields \added{may} lack local detail, limiting its applications to non-smooth and complex registrations. 
        In this challenge, the Fourier-Net employed consists of 5 downsampling convolutional blocks in the contracting path and 1 upsampling convolutional block in the expansive path. Specifically, for an input of size \(H\times W\times D\), the resolution after each block progresses as follows: \([C,D,H,W]\), \([2C, \frac{D}{2}, \frac{H}{2}, \frac{W}{2}]\), \([4C, \frac{D}{4}, \frac{H}{4}, \frac{W}{4}]\), \([8C, \frac{D}{8}, \frac{H}{8}, \frac{W}{8}]\), \([16C, \frac{D}{16}, \frac{H}{16}, \frac{W}{16}]\), and \([3, \frac{D}{8}, \) \(\frac{H}{8}, \frac{W}{8}]\).
        The variable \(C\), representing the number of channels in the first convolutional layer, is set to 16 for this challenge. The Fourier-Net code is available at \url{https://github.com/xi-jia/Fourier-Net}.
        
        A pre-trained model from the ThoraxCBCT dataset was utilised to initialise the weights before the training phase. Once the model was trained, instance-level optimisation was employed for each test pair to refine the displacement estimation.
        Normalised cross-correlation (NCC) and bending energy were used as the similarity measure and regularisation term, respectively. Additionally, a Dice loss on the keypoint masks was incorporated as extra supervision during the network updates.
        
        Keypoint Masks: 
        Instead of using individual key points, each key point was empirically rounded into a \(3\times 3\times 3\) patch, and the Dice loss was computed. This approach was chosen because key points, often labelled by experts or classical methods, may contain some degree of subjectivity or error. Rounding the key points into a patch provides a tolerance for such uncertainty.
        
         Image Augmentation: Online data augmentation was applied during pre-training on the ThoraxCBCT dataset. Specifically, images were randomly rotated, cropped, and subjected to noise addition, using the default probability settings provided by \url{https://github.com/ZFTurbo/volumentations}.
         
        The training process on the private dataset was similar to that used for the ThoraxCBCT dataset, where image similarity, keypoint mask Dice, and bending energy were optimised. Online data augmentation was also employed during this stage.
        The instance-level optimisation involved fine-tuning the trained weights on each test pair. For this challenge, the number of iterations was fixed at 700, with a learning rate of 0.0001 for each test pair.

    \subsection{SSKJLBW: Unsupervised Single-Shot Registration}
        To address the challenges of severe noise interference, large-scale deformations caused by respiratory motion, and limited training data in thoracic cone-beam CT image registration for clinical radiotherapy, a single-shot unsupervised registration framework is proposed. Inheriting the optimisation paradigm of GroupRegNet (\citealp{zhang2021groupregnet}), the method achieves robust registration without annotated data through multi-scale feature fusion and dynamic regularisation constraints. The core workflow begins with noise-adaptive preprocessing of input images, followed by iterative optimisation of deformation fields via a cascaded estimation module, and finally outputs smoothed deformation fields constrained by anatomical masks. 
        
        The framework incorporates physics-guided preprocessing to decouple noise from anatomical deformations. Input HU values are clamped to [0,2048] to suppress metal artefacts and extreme noise, while edge regions are zero-padded to mitigate respiratory motion artefacts. A lightweight 3D UNet processes standardised 128×128×64 image pairs, leveraging hierarchical features for deformation estimation—shallow layers enforce smooth large-scale deformations, while deeper layers refine local details.
        
        Central to the approach is a dynamic template evolution strategy. Initial reference templates are generated by applying random affine transformations to moving images. During iterative alignment, templates are updated by blending historical and newly aligned results, with blending weights gradually shifting from stability-focused (75\% historical) to sensitivity-focused (80\% new) across iterations. This balances global consistency and local precision.
        
        Training leverages anatomy-aware augmentation within torso masks derived from TotalSegmentator (\citealp{wasserthal2023totalsegmentator}), combining spatial transforms (rotation, scaling, translation) with synthetic noise injection. The loss function integrates noise-robust local cross-correlation for structural alignment, spatial gradient penalties for smooth deformations (\citealp{liu2020learning}), and Jacobian regularisation to prevent topological folding.
        
        For inference, a coarse-to-fine hierarchical strategy accelerates computation. Low-resolution (1/8) coarse alignment initialises deformation fields, which are progressively refined at 1/2 and full resolutions. Postprocessing with median filtering removes outlier deformations, achieving clinically feasible 3-minute runtime per case. Experimental validation demonstrates 0.84 DSC for lung alignment under extreme noise ($SNR<5\,dB$), outperforming traditional methods in accuracy and efficiency. The method’s noise-deformation decoupling and adaptive template updating provide a practical solution for limited-data scenarios in thoracic image registration.

    \subsection{SynDeeds} 
        To address the registration challenges posed by low-quality CBCT images, SynDeeds, an enhanced version of Deeds (\citealp{Heinrich12}), is introduced. This variation incorporates a learning-based image synthesis module (\citealp{Chen19}), aiming to enhance the accuracy of registration significantly.

        \begin{figure}[h]
		  \centering
		  \includegraphics[width=1\linewidth]{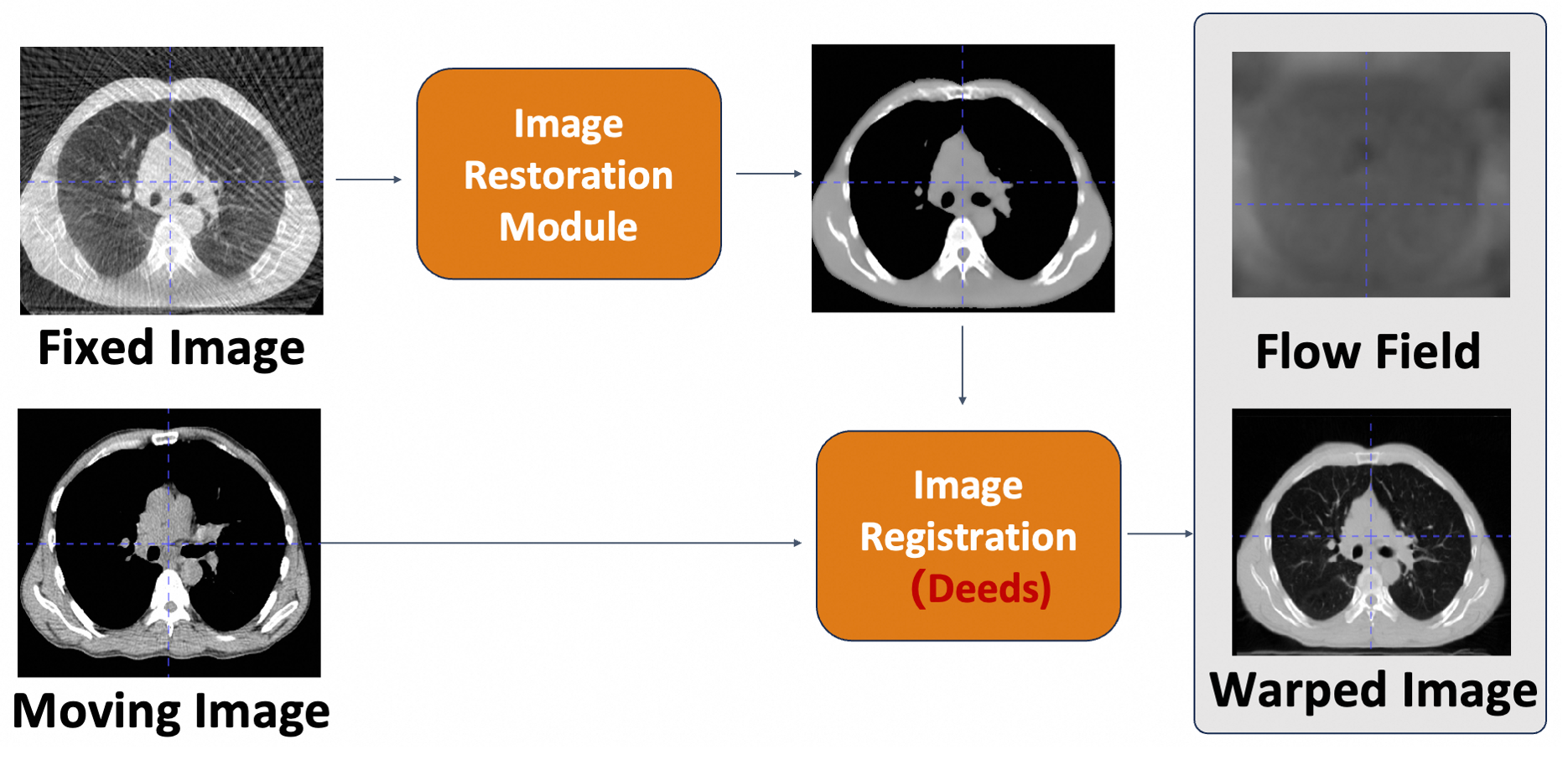}
		  \caption{SynDeeds: Fixed and moving images are registered using Deeds after the fixed image went through the image restoration module.}
          \label{fig:syndeeds}
	    \end{figure}

        The image synthesis approach is framed as an additive restoration problem, defining the clean image as the sum of a degraded image and a variable “noise” component. The loss from network learning is computed as the mean squared error (MSE) between the ground truth residual and the output residual. Therefore, the image synthesis network is trained to learn the residual, instead of reconstructing the clean image directly, as shown in Fig.~\ref{fig:syndeeds}.
        
        \begin{figure}[h]
		  \centering
		  \includegraphics[width=1\linewidth]{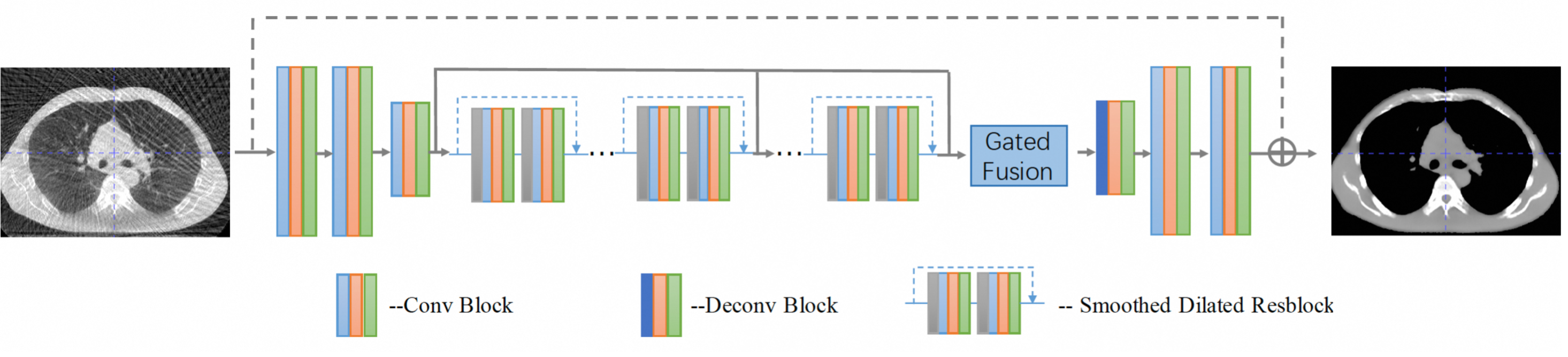}
		  \caption{Overview of the image restoration module.}
	    \end{figure}

    \subsection{TimH: ConstrICON}
        ConstrICON (\citealp{constricon}) is a framework that guarantees inverse consistency by construction. In contrast to other approaches, that utilise penalty terms during training to approximate inverse consistency, ConstrICON achieves this property intrinsically through its architectural design and by parametrising the output transformations of the neural networks by a Lie group. Parametrising the output of a neural network by a Lie group is assumed to be sufficient, since many types of transformations that are used in medical image registration, such as affine or diffeomorphic transforms, are Lie groups.
        
        Given a neural network $N_\theta$ of arbitrary design that takes two input images and whose output can be considered a member of a Lie algebra, a registration network $\Phi$ can be defined, that registers an image $I_A$ to an image $I_B$:
        \begin{align*}
            \Phi[I_A, I_B] = \exp(N_\theta [I_A, I_B] - N_\theta [I_B, I_A]).
            \label{eq:timh1}
        \end{align*}
        The exponential map is implemented using the scaling and squaring method (\citealp{log_euclidean_framework}). The registration network $\Phi$ is inverse consistent by construction, as $\Phi[I_A, I_B] \circ \Phi[I_B, I_A] = Id$, where $Id$ denotes the identity transformation. However, the composition of two or more of these networks is not necessarily inverse consistent. To address this issue, Greer et al. proposed the TwoStepConsistent (TSC) operator (\citealp{constricon}), which ensures inverse consistency when combining two networks.
        
        Let $\sqrt{\Phi[I_A, I_B]}$ be the square root of the output transformation of a registration network, defined as $\sqrt{\Phi[I_A, I_B]} \circ \sqrt{\Phi[I_A, I_B]} = \Phi[I_A, I_B]$, where $\Phi$ is trained to fulfil $I_A \circ \Phi[I_A, I_B] \sim I_B$ and let $\Psi$ be a second registration network, that is trained to register $\hat I_A := I_A \circ \sqrt{\Phi[I_A, I_B]}$ to $\hat I_B := I_B \circ \sqrt{\Phi[I_B, I_A]}$. Then, the TSC operator is defined as 
        \begin{align*}
            TSC \{ \Phi, \Psi \}[I_A, I_B] = \\
            \sqrt{\Phi[I_A, I_B]} \circ \Psi[\hat I_A, \hat I_B] \circ \sqrt{\Phi[I_A, I_B]},
        \end{align*}
        which is inverse consistent if $\Phi$ and $\Psi$ are inverse consistent. Since the TSC operator preserves inverse consistency, it can be recursively applied, thereby extending the procedure to an N-step registration approach that remains inverse consistent.
        
        For the ThoraxCBCT task, three residual U-Nets (\citealp{8309343}) are employed and a four-step preprocessing pipeline is applied. First, images and their corresponding masks were affinely aligned. Then, empty slices were removed, followed by histogram matching. Finally, to reduce GPU memory usage and computational load, images were downscaled to $128 \times 128 \times 128$ voxels, approximately $47.1\%$ of the original preprocessed size.
        The square roots of the output transformations of the registration networks were learned implicitly, i.e., we trained an inverse consistent registration network $\tilde \Phi$ \added{to fulfil} $I_A \circ \tilde \Phi [I_A, I_B] \circ \tilde \Phi [I_A, I_B] \sim I_B$ instead of calculating the square root of a network's output that directly registers $I_A$ to $I_B$. For each registration network, the output of $N_\theta[I_A, I_B] - N_\theta [I_B, I_A]$ was regularised using the selective Jacobian determinant regularisation loss (\citealp{fast_symmetric_diff}). The final output of the combined registration networks, was regularised using bending energy loss (\citealp{10.1007/3-540-45729-1_33}), with local normalised cross-correlation (LNCC) as the similarity metric. Additionally, it was found that further smoothing the stationary velocity field (SVF) reduced the inverse consistency error, so average pooling with a window size of 3 twice was applied. During inference, the resulting displacement fields were upscaled to the original image resolution.

    \subsection{VoxelMorph++}
        \added{VoxelMorph++ (\citealp{heinrich2022voxelmorph++}) extends VoxelMorph (\citealp{Balakrishnan19}) by introducing heatmap-based displacement prediction with keypoint-based self-supervision and instance optimisation. Instead of directly regressing dense displacement fields, the method predicts probabilistic displacement heatmaps, enabling more robust handling of large deformations and ambiguous optimisation landscapes.}

        \added{The framework uses a U-Net backbone to extract shared image features and predicts discretised displacement distributions supervised by automatically generated keypoint correspondences. For the OncoReg task, Förstner keypoints were extracted in the moving image and weak correspondences were obtained using \textit{ConvexAdam} (\citealp{siebert2021fast}). Trunk masks were used to restrict the region of interest, and affine augmentation was applied during training.}
        
       \added{Registration is driven by a non-local MIND loss, while sparse displacements are converted into dense deformation fields using thin-plate splines. Furthermore, an instance optimisation step refines the network prediction using MIND-based optimisation and regularisation. An ablation study demonstrated that this refinement is essential for achieving strong performance on the OncoReg task.}
        
        \added{By combining heatmap-based displacement prediction, weakly supervised keypoint correspondences, MIND similarity, and instance optimisation, \textit{VoxelMorph++} achieves state-of-the-art performance for large-deformation medical image registration.}

\section{Results}
    \subsection{ThoraxCBCT}
\begin{figure*}[h]
        		\centering
        		\includegraphics[width=0.8\linewidth]{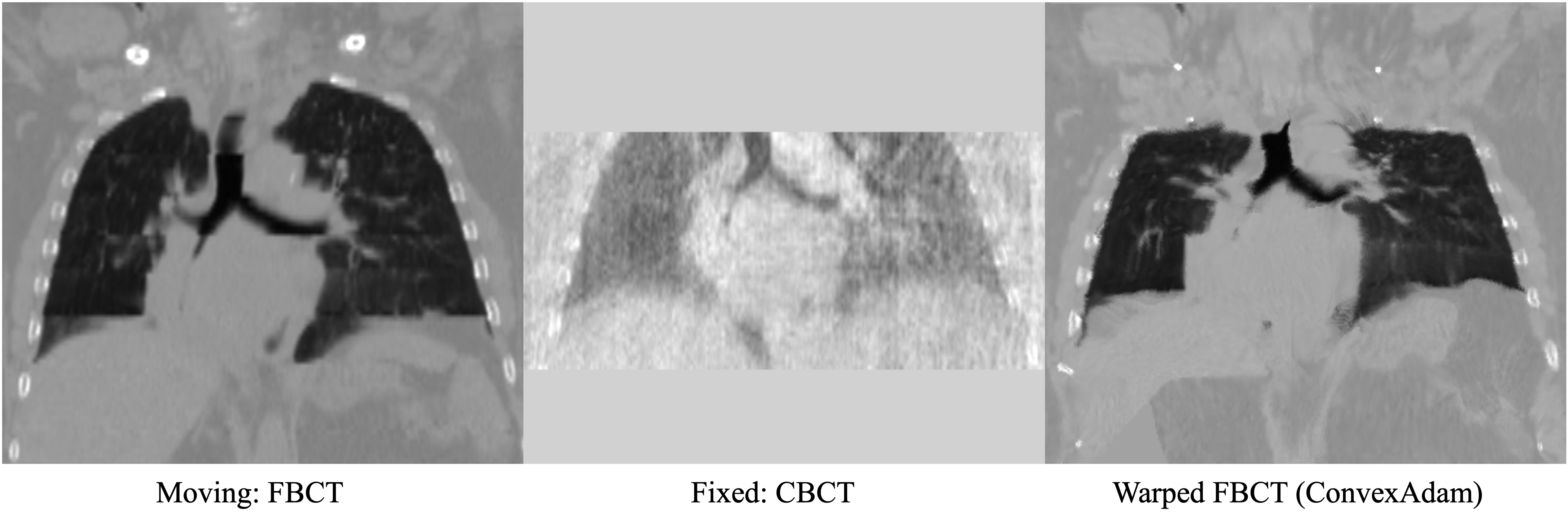}
        		\caption{Exemplary ThoraxCBCT image pair before registration (left: moving FBCT; middle: fixed CBCT) and after registration with ConvexAdam (right: warped FBCT).}
                \label{fig:warped}
        	\end{figure*}

        The quantitative results for the ThoraxCBCT task are summarised in Table~\ref{tab:thoraxcbct}. While \textit{ConvexAdam} achieved the highest overall ranking across all evaluation metrics, the official winner of the task was \textit{SynDeeds}. Since \textit{ConvexAdam} was provided by the organi\added{s}ers as a baseline method, it was not eligible for ranking. However, the performance of \textit{ConvexAdam} and \textit{SynDeeds} was very close across most metrics. An exemplary image pair is shown together with the warped moving image based on a displacement field achieved with \textit{ConvexAdam} \deleted{is shown} in Fig.~\ref{fig:warped}.
        
        When examining individual performance measures, \textit{SynDeeds} achieved the highest mean Dice \added{coefficient} (DSC), while \textit{Fourier-Net} obtained the lowest Target Registration Error (TRE). The best Hausdorff Distance 95th percentile (HD95) was observed for \textit{SSKJLBW}, although this method exhibited a less smooth displacement field.
        
\begin{figure*}[t]
    		\centering
    		\includegraphics[width=0.8\linewidth]{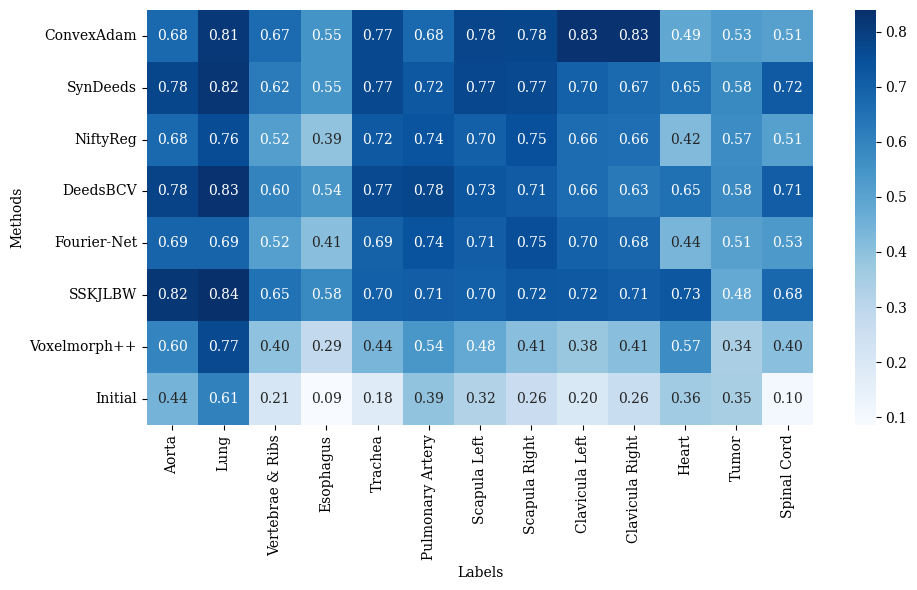}
    		\caption{ThoraxCBCT: Dice \added{c}oefficients for each anatomical label and registration method, averaged across all test images.}
            \label{fig:thoraxcbct_dice}
    	\end{figure*}

        A more detailed breakdown of Dice coefficients achieved by all methods across different anatomical labels is shown in Fig.~\ref{fig:thoraxcbct_dice}. \added{Across all methods, the lung consistently achieved the highest Dice scores, reflecting the comparatively large size and well-defined boundaries of this structure. In contrast, smaller and more elongated structures such as the oesophagus and spinal cord proved substantially more challenging. \textit{ConvexAdam} demonstrated particularly consistent performance across almost all anatomical labels, achieving among the highest Dice scores for both bony and soft-tissue structures. \textit{SynDeeds} achieved similarly strong performance and showed notable improvements over the related \textit{DeedsBCV} for several structures.}
        
        \added{The TRE distributions in Fig.~\ref{fig:thoraxcbct_tre} reveal a somewhat different ranking of methods. While \textit{Fourier-Net} achieved the lowest mean TRE, this advantage did not translate into the highest Dice scores. Conversely, \textit{ConvexAdam} and \textit{SynDeeds} obtained some of the strongest segmentation scores despite slightly higher TRE values. \textit{VoxelMorph++} exhibited substantially larger registration errors and greater variability than the leading methods, consistent with its comparatively lower Dice scores across most anatomical structures.}
        
    \subsection{OncoReg}
        
        For the initial challenge evaluation, we used a standard train-validation-test split. However, since many submitted methods did not rely on training data, we later expanded our evaluation to be more robust and meaningful. After completing additional annotations for all images, we conducted a three-fold cross-validation for all trainable models and evaluated the non-trainable methods on all additionally annotated images to ensure a fair and comprehensive comparison.
        
\begin{figure*}[h]
        \centering
        \begin{minipage}[t]{0.49\linewidth}
            \centering
            \includegraphics[width=\linewidth]{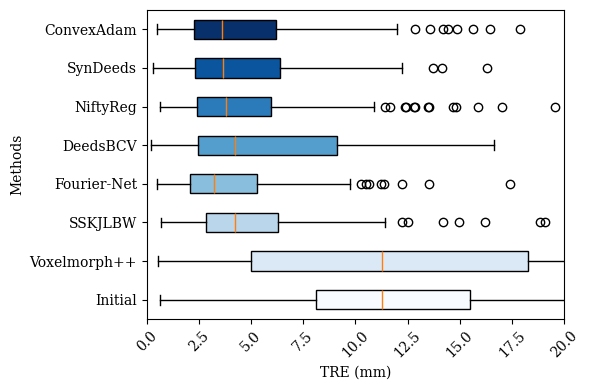}
            \caption{ThoraxCBCT: Target Registration Error averaged over all landmarks and test images for each registration method.}
            \label{fig:thoraxcbct_tre}
        \end{minipage}
        \hfill
        \begin{minipage}[t]{0.49\linewidth}
            \centering
            \includegraphics[width=\linewidth]{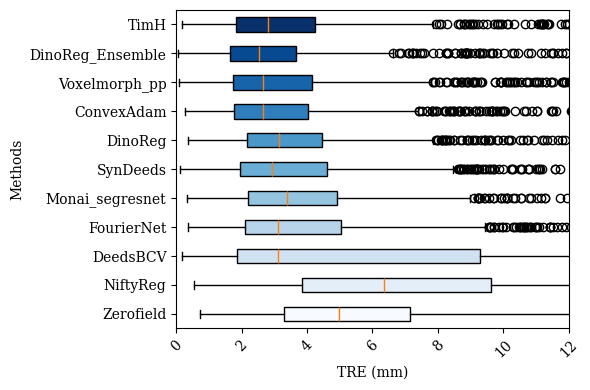}
            \caption{OncoReg: Target Registration Error averaged over all landmarks and \deleted{test} images for each registration method.}
            \label{fig:oncoreg_tre}
        \end{minipage}
        \end{figure*}

        All trainable models were trained and evaluated on an NVIDIA RTX 4000 SFF Ada Generation GPU with 20 GB of memory. Importantly, runtime was not considered in the ranking. \deleted{The quantitative results are summarised in Table~\ref{tab:oncoreg} and visualised in Fig.~\ref{fig:oncoreg_tre} and Fig.~\ref{fig:oncoreg_dice}. More detailed results for bony structures (ribs and vertebrae) can be found in Appendix~\ref{AppA}.}
        
        The official winner of the challenge was the method \textit{DINO-Reg Ensemble}, which achieved the lowest target registration error and the highest Dice coefficient among all submissions. However, a late submission (\textit{TimH}), which was not included in the initial ranking, outperformed \textit{DINO-Reg Ensemble} in terms of the 95th percentile of the Hausdorff distance and the 70th percentile of the TRE, ultimately achieving the best overall performance.
        
        These two leading methods were closely followed by \textit{VoxelMorph++} and \textit{ConvexAdam}, while \textit{DeedsBCV} and \textit{NiftyReg} ranked lowest overall.
        
        \added{The quantitative results are summarised in Table~\ref{tab:oncoreg} and visualised in Fig.~\ref{fig:oncoreg_tre} and Fig.~\ref{fig:oncoreg_dice}. More detailed results for bony structures (ribs and vertebrae) can be found in Appendix~\ref{AppA}. }
        
        \added{A detailed analysis of the Dice scores across anatomical structures reveals substantial differences in registration difficulty (see Fig.~\ref{fig:oncoreg_dice}). Large structures such as the lungs consistently achieved the highest Dice scores across all methods, with \textit{TimH} reaching a DSC of 0.94. In contrast, smaller and anatomically more complex structures, including the scapulae and tumours, remained considerably more challenging. Nevertheless, all top-performing methods achieved substantial improvements over the initial alignment for every evaluated structure.
        The strongest overall segmentation performance was observed for \textit{TimH}. \textit{DINO-Reg Ensemble} demonstrated similarly strong performance and showed particularly robust behaviour across all structures. \textit{VoxelMorph++} achieved competitive results despite relying on a considerably different registration paradigm and consistently ranked among the top-performing methods for both bony and soft-tissue anatomy.}
        
        \added{The TRE distributions shown in Fig.~\ref{fig:oncoreg_tre} reveal a slightly different ranking. While \textit{TimH} and \textit{DINO-Reg Ensemble} achieved the lowest median registration errors, \textit{ConvexAdam} produced similarly low TRE values despite obtaining lower Dice scores than the leading methods. }
        
        \added{The vertebral and rib-specific analyses presented in Appendix~\ref{AppA} demonstrate that larger improvements over the initial alignment were obtained for the ribs than for the vertebrae. \textit{DINO-Reg Ensemble, ConvexAdam}, and \textit{VoxelMorph++} consistently achieved high Dice scores across almost all vertebral levels and ribs, whereas \textit{TimH} remained substantially below the performance of the leading approaches for the vertebrae. Interestingly, performance generally decreased towards the outermost ribs and upper cervical vertebrae, being especially prominent for the upper and lower vertebrae with the method \textit{TimH}.}
        
        \added{The relationship between landmark-based and segmenta-tion-based evaluation metrics is investigated by comparing the mean TRE and mean DSC achieved by each method in Fig.~\ref{fig:dsc_tre}. As expected, methods with lower TRE generally obtain higher DSC values, resulting in a moderate negative correlation ($\rho=-0.52$). 
        Fig.~\ref{fig:dsc_hd} visualises the relationship between DSC and HD95 for five clinically relevant anatomical structures.} The lungs consistently exhibit high Dice scores, with the expected inverse relationship to HD95. In contrast, smaller structures such as the scapulae yield lower Dice scores. These patterns are also evident in Fig\added{.}\deleted{ure}~\ref{fig:oncoreg_dice}.
        Tumour registration reveals three distinct trends: high Dice scores are consistently associated with low HD95, but among the lower Dice scores, both low and very high HD95 values are observed. This behaviour differs from that seen in other anatomical structures.
        
        To ensure transparency in our evaluation process, we provide the complete code for evaluation and ranking on GitHub\footnote{https://github.com/MDL-UzL/L2R/}.

        \begin{table*}[h] 
    		\centering
    		\caption{Quantitative \added{r}esults of the ThoraxCBCT task.}
    		\begin{tabular}{lccccccc}
    			& \textbf{TRE\(\downarrow\)}$\textcolor{white}{a}_{mm}$ & \textbf{TRE30\(\downarrow\)}$\textcolor{white}{a}_{mm}$ & \textbf{DSC\(\uparrow\)}$\textcolor{white}{a}_\%$ & \textbf{HD95\(\downarrow\)}$\textcolor{white}{a}_{mm}$ & \textbf{SDlogJ\(\downarrow\)} & \textbf{Rank\(\uparrow\)}\\
    			\hline
                Initial & 11.91 & 16.26 & 29.04 & 48.13 &  &  \\
                \hline
                ConvexAdam & 4.93 & 4.98 & 68.40 & 38.00 & 0.10 & 0.710 \\
                SynDeeds & 6.01 & 5.67 & \textbf{70.16} & 22.97 & 0.15 & 0.705 \\
                NiftyReg & 4.90 & 4.51 & 62.11 & 43.41 & \textbf{0.05} & 0.577 \\
                DeedsBCV & 8.76 & 7.38 & 69.00 & 22.38 & 0.16 & 0.537 \\
                Fourier-Net & \textbf{4.03} & \textbf{3.85} & 61.91 & 47.80 & 0.07 & 0.535 \\
                SSKJLBW & 5.80 & 7.44 & 69.45 & \textbf{17.60} & 0.26 & 0.515 \\
                VoxelMorph++ & 13.06 & 16.75 & 46.28 & 33.34 & 0.12 & 0.357 \\
    		\end{tabular}
            \label{tab:thoraxcbct}
    	\end{table*}
    
    \begin{table*}[h] 
		\centering
		\caption{Quantitative Results of the OncoReg task. \added{Runtime measurements were only recorded for submissions evaluated through the original challenge pipeline. For methods submitted after the official challenge evaluation, runtime information is not available.}}
		\begin{tabular}{lcccccccc}
			& \textbf{TRE\(\downarrow\)}$\textcolor{white}{a}_{mm}$ & \textbf{TRE30\(\downarrow\)}$\textcolor{white}{a}_{mm}$ & \textbf{DSC\(\uparrow\)}$\textcolor{white}{a}_\%$ & \textbf{HD95\(\downarrow\)}$\textcolor{white}{a}_{mm}$ & \textbf{SDlogJ\(\downarrow\)} & \textbf{RT\(\downarrow\)}$\textcolor{white}{a}_s$ & \textbf{Rank\(\uparrow\)}\\
			\hline
            Initial & 6.41 & 11.43 & 43.38 & 32.68 &  &  &  \\
            \hline
            TimH & 5.30 & \textbf{6.31} & 63.21 & \textbf{6.92} & \textbf{0.039} & \added{N/A} & 0.73 \\
            DINO-Reg Ens. & \textbf{3.94} & 7.63 & \textbf{63.89} & 31.12 & 0.062 & $>300$ & 0.67 \\
            VoxelMorph++ & 4.66 & 8.45 & 62.36 & 26.86 & 0.081 & $<60$ & 0.60 \\
            ConvexAdam & 4.16 & 7.89 & 60.13 & 31.03 & 0.066 & $(<5)$ & 0.60 \\
            DINO-Reg & 4.53 & 8.46 & 57.64 & 81.83 & \textbf{0.039} & $>300$ & 0.57 \\
            SynDeeds & 4.51 & 7.93 & 59.93 & 28.29 & 0.100 & \added{N/A} & 0.52 \\
            MONAI SegResNet & 4.78 & 8.97 & 53.63 & 32.37 & 0.048 & \added{N/A} & 0.46 \\
            Fourier-Net & 6.01 & 10.52 & 60.67 & 23.53 & 0.103 & $>300$ & 0.44 \\
            DeedsBCV & 7.99 & 12.28 & 58.33 & 22.33 & 0.154 & $<15$ & 0.34 \\
            NiftyReg & 7.98 & 12.38 & 37.30 & 32.71 & 0.061 & $<60$ & 0.31 \\
           \label{tab:oncoreg}
		\end{tabular}
	\end{table*}

	\begin{figure*}[t]
		\centering
		\includegraphics[width=0.8\linewidth]{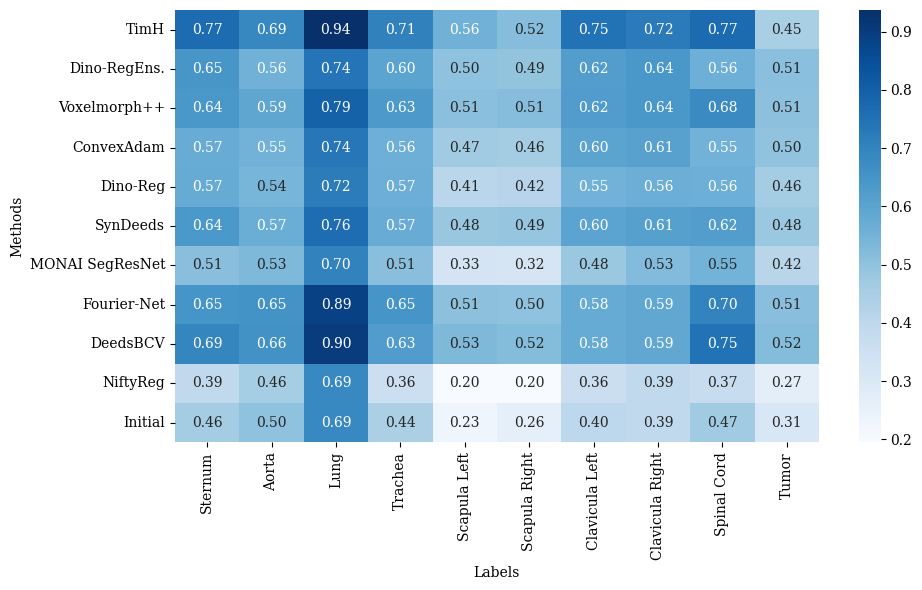}
		\caption{OncoReg: Dice \added{c}oefficients for ten anatomical labels and each registration method, averaged across all \deleted{test} images.}
        \label{fig:oncoreg_dice}
	\end{figure*}

\newpage
\section{Discussion}

    \subsection{Evaluation Metrics and Annotation Challenges}
        Evaluating image registration remains a complex task, as standard metrics often fail to capture all aspects of registration quality and are highly dependent on annotation accuracy. In this challenge, the primary evaluation metric was the Target Registration Error, which relies on manually placed anatomical landmarks. However, landmark positioning is inherently uncertain in certain structures; while the carina served as a reliable reference point, the aortic arch and vertebral midpoints were more challenging to pinpoint precisely. To minimise inter-observer variability, all landmarks within a given image pair were annotated by the same medical expert, ensuring internal consistency.
        Significant effort was dedicated to refining these annotations, including extending the publicly available 4D-Lung dataset and comprehensively annotating our private dataset. This investment considerably enhanced the dataset’s usability. However, annotation challenges were particularly pronounced in cone-beam CT images, where lower resolution, poor contrast, and strong artefacts complicated precise landmark placement. In these cases, tools like TotalSegmentator proved invaluable, significantly reducing annotation time while maintaining quality in a semi-automated setting.
        Despite the comprehensive measures taken, the analysis of the registration results revealed that the label masks, particularly on the artefact-prone CBCT scans, often exhibited poorly smoothed surfaces. While the masks generally covered the approximate volume of the structures, they tended to be less reliable at the boundaries. This particularly affected the Dice coefficient and surface distance for small or thin structures, such as the clavicles and ribs.
        Furthermore, varying fields of view, as illustrated in Fig.~\ref{fig:warped}, introduced an additional challenge for registration. As the CBCTs cover a more limited anatomical region compared to the FBCTs, which often include more extensive surrounding anatomy, many methods struggled to accurately register structures located near the image borders. This difficulty is reflected in the OncoReg task by the low Dice scores for the scapulae, lumbar and cervical vertebrae, as well as the most inferior and superior ribs, all of which are typically situated near the edges of the images. This trend is particularly evident in Fig.~\ref{fig:oncoreg_vertebrae} and Fig.~\ref{fig:oncoreg_ribs}.
        Sample-based inspection further revealed that the notably high variability in Hausdorff distances was largely attributable to distortions at the image margins. The method \textit{TimH} addresses this problem through extensive preprocessing: FBCT images are cropped to match the spatial extent of the corresponding CBCTs, ensuring alignment in the regions where both modalities contain anatomical information. This approach leads to significantly lower Hausdorff distances (see Table~\ref{tab:oncoreg}). These insights are of high relevance for data preprocessing in future challenges.
        
        \added{The challenge ranking design places greater emphasis on registration accuracy than on deformation-field regularity. This design choice was motivated by the primary clinical objective of the challenge, namely accurate alignment of tumours and organs-at-risk in the context of image-guided radiotherapy. Metrics such as TRE, DSC, and HD95 directly reflect the accuracy of anatomical correspondence and are therefore most closely related to the intended clinical application. Nevertheless, deformation plausibility remains an important consideration, particularly for applications requiring anatomically realistic transformations. While SDlogJ was included to assess deformation smoothness, it captures only a limited aspect of transformation quality. 
        We note that methods with stronger regularisation strategies, such as \textit{TimH}, performed competitively in the plausibility metric, highlighting the value of such approaches. Future challenge editions could investigate alternative ranking schemes or additional deformation-quality metrics to further balance registration accuracy and transformation plausibility.}
        
        \added{Taking a closer look at the accuracy metrics: the finding of a moderate negative Spearman correlation of $\rho=-0.52$ between DSC and TRE values indicates that landmark localisation accuracy and segmentation overlap capture different aspects of registration quality. This observation supports the use of multiple complementary evaluation metrics, as improvements in one metric do not necessarily translate directly into improvements in another. In particular, some methods achieve comparable DSC values despite noticeable differences in TRE, suggesting that anatomical overlap and landmark correspondence provide distinct information about registration performance. A similar observation can be made when comparing DSC and HD95 values across anatomical structures. While lower HD95 values generally correspond to higher DSC scores, the relationship is strongly structure-dependent. Large structures such as the lung maintained consistently high DSC values even at moderate HD95 values, whereas smaller and more elongated structures, particularly the tumour and trachea, exhibited substantially greater sensitivity to boundary deviations. While these findings are not particularly new, they further highlight that overlap- and distance-based metrics capture complementary aspects of registration quality and should therefore both be considered when evaluating deformable registration methods. Further, evaluating multiple metrics decreases the unavoidable human bias and errors introduced by manual annotation. }
        
        \added{As the last metric considered in the evaluation, inference runtime did not contribute to the final challenge ranking but it constitutes an important factor for the clinical deployment of deformable registration methods, particularly in adaptive radiotherapy workflows. Several of the best-performing methods in this challenge relied on instance-level optimisation and therefore exhibited registration runtimes of several minutes per image pair. While such runtimes may be acceptable for offline analysis or treatment replanning scenarios, they remain challenging for time-critical online applications. At the same time, many registration methods currently used in clinical practice are still based on iterative optimisation and are subject to similar computational limitations. Since the primary aim of OncoReg was to benchmark registration accuracy on a challenging FBCT-to-CBCT task, runtime was not included in the final ranking. However, as deformable image registration moves closer to routine clinical adoption in radiotherapy and increasingly replaces rigid or manual registration workflows, computational efficiency will become an increasingly important consideration and should receive greater attention in future challenges.}

    \subsection{Challenge Design and Implementation}
        \added{A limitation of both challenge phases is the relatively small number of available cases. For the ThoraxCBCT task, the dataset included all publicly available FBCT-CBCT pairs of the 4D-Lung TCIA dataset with sufficient image quality and complete data required for evaluation. Expanding the dataset further would have introduced a domain shift and more importantly required substantial additional annotation efforts. Consequently, the challenge reflects the current reality of CBCT-based radiotherapy datasets, where large, comprehensively annotated public benchmarks remain scarce.}
        
        \added{The fixed train-validation-test split used in ThoraxCBCT was chosen to align with the established Learn2Reg challenge framework and to ensure a consistent benchmark across all participating methods. Rather than selecting cases randomly, the split was designed to represent the variability present in the dataset. Nevertheless, performance estimates obtained from a small test cohort are inherently associated with increased uncertainty and may be influenced by the specific composition of the test set.}
        
        \added{To mitigate this limitation, the second phase of the challenge (OncoReg) was re-evaluated using three-fold cross-validation with annotations available for the complete dataset. While the dataset size remains limited, this additional evaluation provides a more robust assessment of method performance and reduces the influence of individual cases on the final ranking. Future challenge editions would benefit from larger multi-centre datasets, enabling both more extensive cross-validation and a broader assessment of algorithm robustness across diverse patient populations and acquisition settings.}
        
        \added{The use of planning FBCT scans acquired at maximum inspiration and treatment CBCT scans acquired at expiration was chosen to reflect a realistic challenging radiotherapy scenario. In many clinical settings, particularly those without dedicated 4D imaging or respiratory-gating capabilities, phase-matched acquisitions are not routinely available. Consequently, registration algorithms must simultaneously address modality differences and substantial respiratory motion. While this increases the difficulty of the task, it better reflects challenges encountered in clinical practice. Nevertheless, a complementary inspiration-to-inspiration benchmark could be valuable in future challenge editions, as it would allow evaluation of registration performance with reduced respiratory motion and facilitate a more isolated assessment of modality-related registration difficulties.}
        
        Participants had the option to submit results via an automatic evaluation platform on grand-challenge.org for the ThoraxCBCT task, but this feature was rarely utilised. Instead, most teams preferred local evaluation. \added{For the OncoReg part}, the provided \added{aid in form of a }container example was widely adopted, particularly for handling data loading and saving, which streamlined debugging in the Type 3 challenge setup. Despite these advantages, some recurring issues\added{, such as an insufficiently clear communication of the limits of computational resources and inflexibility in the evaluation pipelines leading to time-consuming manual debugging,} were identified that \added{may} be addressed in future Type 3 challenges.
        
        Organi\added{s}ing such a challenge demands substantial time and resources, but efficient communication and accessibility between organisers and participants played a crucial role in ensuring a smooth workflow. Future iterations should build on this experience to optimise logistical aspects and further reduce participation barriers.
        
    \subsection{Key Findings and Methodological Insights} 
        A key takeaway from the challenge is that the \added{highest ranked} registration methods leveraged strong feature extraction strategies. The top-performing methods, \textit{ConvexAdam} and \textit{DINO-Reg}, demonstrate two distinct but effective approaches:
        
        \begin{itemize}
            \item \textbf{ConvexAdam} combines handcrafted and semantic features within a two-stage optimisation process. The extracted semantic features play a crucial role in achieving robust registration across different modalities.
            \item \textbf{DINO-Reg} employs self-supervised learning with Vision Transformers (ViTs), using deep feature representations from a DINO-pretrained ViT to enhance registration performance across modalities.
        \end{itemize}
        
        Both methods prioritise feature extraction but differ in execution, \textit{ConvexAdam} explicitly designs features, whereas \textit{DINO-Reg} learns them in a self-supervised manner.
        Beyond these top approaches, an intriguing finding was the potential for further improving \textit{ConvexAdam}’s performance by integrating it with \textit{VoxelMorph++}. Specifically, using \textit{ConvexAdam}-derived Förstner keypoints as input for \textit{VoxelMorph++} led to a refined registration process. This effect was particularly evident in OncoReg, where preprocessed keypoints significantly improved \textit{VoxelMorph++'s} accuracy. However, in ThoraxCBCT, \textit{VoxelMorph++} performed notably worse than other methods. Further analysis revealed that its performance was hampered by suboptimal keypoint registration, which proved particularly detrimental in the weakly supervised setting. A post-challenge re-evaluation using keypoints registered by ConvexAdam showed substantial improvement\added{,} although this did not affect the official rankings.
        Another notable result was the enhancement of \textit{DeedsBCV} by Li et al., leading to the superior performance of \textit{SynDeeds} in both registration tasks. This demonstrates that traditional methods, when effectively adapted, can still compete with or even outperform newer deep-learning-based approaches.
        
        \added{The per-structure analysis further provides insight into how different registration strategies interact with anatomical characteristics. Across both challenge tasks, large structures such as the lungs consistently achieved the highest Dice scores, regardless of the underlying registration method. Their large spatial extent and rich anatomical context provide abundant image information for both feature-based and learning-based approaches. In contrast, smaller and elongated structures, including the oesophagus, spinal cord, and tumour regions, remained considerably more challenging. For these structures, even small local registration errors can lead to substantial reductions in overlap metrics.}
        
        \added{Interestingly, methods based on robust feature representations, such as \textit{ConvexAdam} and \textit{DINO-Reg}, tended to maintain more consistent performance across different anatomical regions than methods relying primarily on local optimisation. This suggests that strong feature extraction improves not only overall registration accuracy but also robustness to anatomical variability and local image artefacts.}
        
        \added{An additional insight emerged from the strong performance of the late submission \textit{TimH}. Unlike most participating methods, \textit{TimH} explicitly addressed the mismatch in field-of-view between FBCT and CBCT scans by cropping the planning CT to the anatomical region covered by the corresponding CBCT. This preprocessing step substantially reduced distortions in areas where no CBCT image information was available and was particularly beneficial for boundary-based metrics such as HD95. The results suggest that a considerable portion of the registration challenge arises not only from modality differences and respiratory motion but also from the incomplete anatomical overlap between the two image acquisitions.
        This observation is further supported by the behaviour of classical optimisation-based methods such as \textit{NiftyReg}. Visual inspection of the resulting deformation fields on OncoReg revealed strong distortions in image regions outside the CBCT field-of-view, indicating that the optimisation was partially driven by anatomy that was not represented in both images. However, the comparatively stronger performance of \textit{NiftyReg} on ThoraxCBCT suggests that the limited field-of-view alone does not fully explain the observed performance differences. A more likely explanation is the presence of a domain shift between the ThoraxCBCT and OncoReg datasets, including differences in scanners, image appearance, artefacts, and acquisition characteristics. Since the challenge implementation largely relied on hyperparameters and preprocessing strategies originally developed for ThoraxCBCT, the method may not have been optimally adapted to the OncoReg dataset. It is therefore plausible that improved preprocessing, explicit field-of-view handling, or a more thorough re-optimisation of hyperparameters could have yielded substantially better results.}
        
        \added{More generally, these findings highlight that preprocessing choices, domain adaptation, and the explicit handling of partial image overlap can have an impact comparable to algorithmic innovations and should receive greater attention in future benchmark studies. This observation suggests that future registration benchmarks should not only compare registration architectures, but also explicitly evaluate preprocessing strategies and robustness to domain shift, as these factors may substantially influence performance rankings.}
        
        \added{Considering the computational efficiency of the submitted methods, several top-performing approaches, such as \textit{DINO-Reg Ensemble} and \textit{Fourier-Net}, relied on computationally expensive instance-level optimisation with runtimes exceeding five minutes per image pair, \textit{ConvexAdam} achieved highly competitive performance with runtimes below five seconds. Despite this substantial difference in computational cost, \textit{ConvexAdam} remained among the strongest methods across both tasks. This suggests that the quality of the extracted image features may be at least as important as extensive optimisation procedures. In particular, the results indicate that strong feature representations can compensate for reduced optimisation effort and may enable registration methods that combine high accuracy with clinically practical runtimes. As deformable image registration moves closer to routine clinical deployment, this trade-off between feature quality, registration accuracy, and computational efficiency is likely to become increasingly important.}
        
    \subsection{Conclusion and Future Directions}
        \added{The challenge provides a comprehensive benchmark for FBCT-to-CBCT registration in radiotherapy and offers several insights into the current state of deformable image registration. Across both challenge phases, the strongest methods consistently relied on robust feature representations, highlighting the central role of feature extraction in modern registration pipelines.}
        
        \begin{itemize}
            \item \added{Beyond algorithmic design, the challenge revealed that \textbf{registration performance is strongly influenced by dataset characteristics, preprocessing strategies, and adaptation to the specific clinical setting}. The success of \textit{TimH} and the observed limitations of several methods in regions with incomplete field-of-view demonstrate that handling partial anatomical overlap can be as important as the choice of registration architecture itself. Similarly, the differing behaviour of some methods between ThoraxCBCT and OncoReg highlights the impact of domain shift and the importance of dataset-specific adaptation.}
            \item \textbf{Feature extraction is critical}: Whether through explicitly designed features (ConvexAdam) or learned representations (\textit{DINO-Reg}), robust feature extraction remains a cornerstone of successful registration. Furthermore, combining classical registration techniques with deep learning shows to be promising.
            \item \added{The challenge further illustrates that no single evaluation metric is sufficient to fully characterise registration quality. Landmark-based, overlap-based, and surface-based metrics were found to capture complementary aspects of performance, supporting the \textbf{use of multi-metric evaluation protocols} for future registration benchmarks.}
            \item \added{Finally, while the highest-performing methods achieved impressive registration accuracy, several relied on computationally expensive optimisation procedures. As deformable image registration moves closer to routine clinical deployment, future research should increasingly focus on \textbf{balancing registration accuracy, robustness, computational efficiency, and generalisability} across institutions and imaging protocols.}
        \end{itemize}
        
        \added{Future iterations of the challenge would benefit from larger multi-centre datasets, more extensive cross-validation, improved annotation pipelines, and additional evaluation criteria that better reflect clinical applicability. We hope that the datasets, benchmark infrastructure, and findings presented in this work will support the continued development of robust and clinically relevant deformable image registration methods for image-guided radiotherapy.}


\acks{This work was partly funded by the German Federal Ministry of \added{Research, Technology and Space (BMFTR)} (grant number 01KD2212A).}

%
\ethics{The work follows appropriate ethical standards in conducting research and writing the manuscript, following all applicable laws and regulations regarding treatment of animals or human subjects.}

\coi{We declare we do not have conflicts of interest.}

\data{The ThoraxCBCT data is available via \href{https://learn2reg.grand-challenge.org/oncoreg/}{learn2reg.grand-challenge.org/oncoreg/}. As described above, the OncoReg dataset cannot be made publicly available. However, requests for evaluation on this dataset are welcome; evaluations will be performed in-house by the organisers.}

\bibliography{bib_oncoreg}


\appendix
\section{Additional visualisations for evaluation}
    \begin{figure}[h]
        \includegraphics[width=\linewidth]{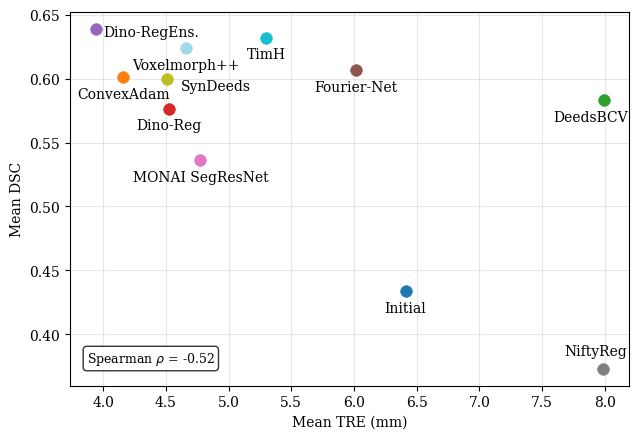}
        \caption{OncoReg: \added{Relationship between mean Target Registration Error and mean Dice Similarity Coefficient across all registration methods. Each point represents one method, with TRE averaged across all landmark correspondences and DSC averaged across all images and anatomical structures. A moderate negative correlation (Spearman $\rho=-0.52$) is observed.}}
        \label{fig:dsc_tre}
    \end{figure}
    \begin{figure}[h]
        \includegraphics[width=\linewidth]{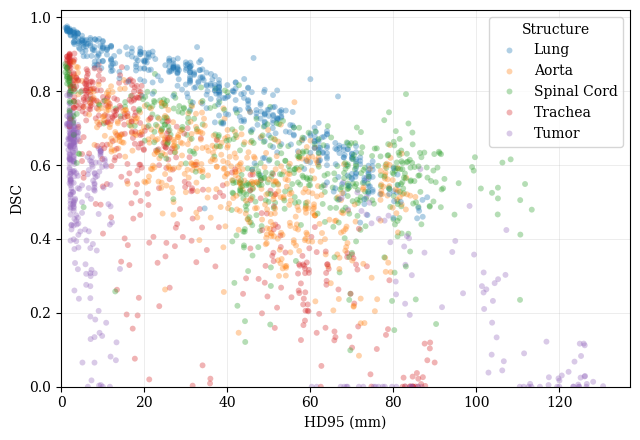}
        \caption{OncoReg: \added{Relationship between Dice Similarity Coefficient and 95th percentile Hausdorff Distance for five clinically relevant anatomical structures across all images and registration methods. }}
        \label{fig:dsc_hd}
\end{figure}
\label{AppA}

    	\begin{figure*}[h]
		\centering
		\includegraphics[width=0.8\linewidth]{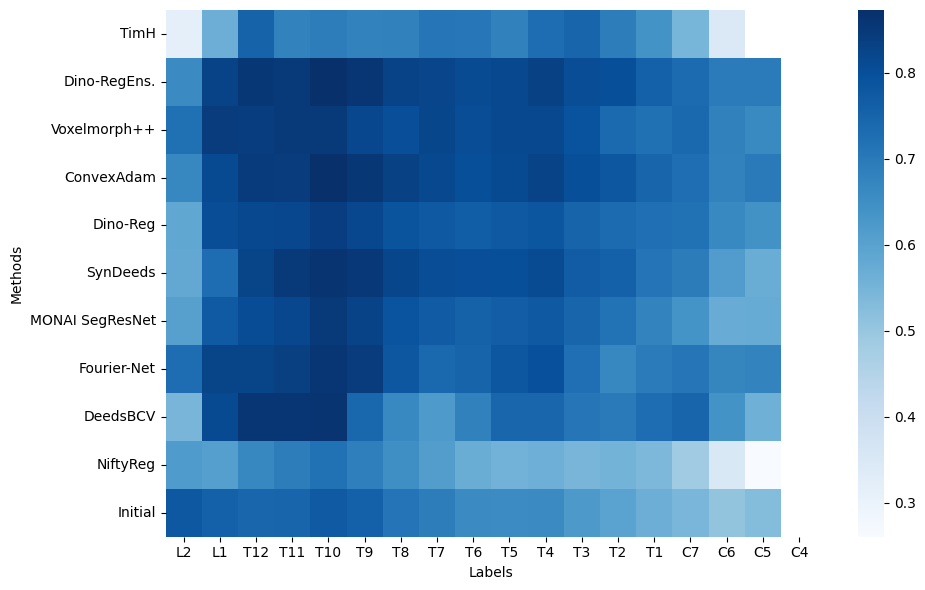}
		\caption{OncoReg: Dice \added{c}oefficients for individual vertebrae-labels and each registration method, averaged across all \deleted{test} images.}
        \label{fig:oncoreg_vertebrae}
	\end{figure*}

    	\begin{figure*}[h]
		\centering
		\includegraphics[width=0.8\linewidth]{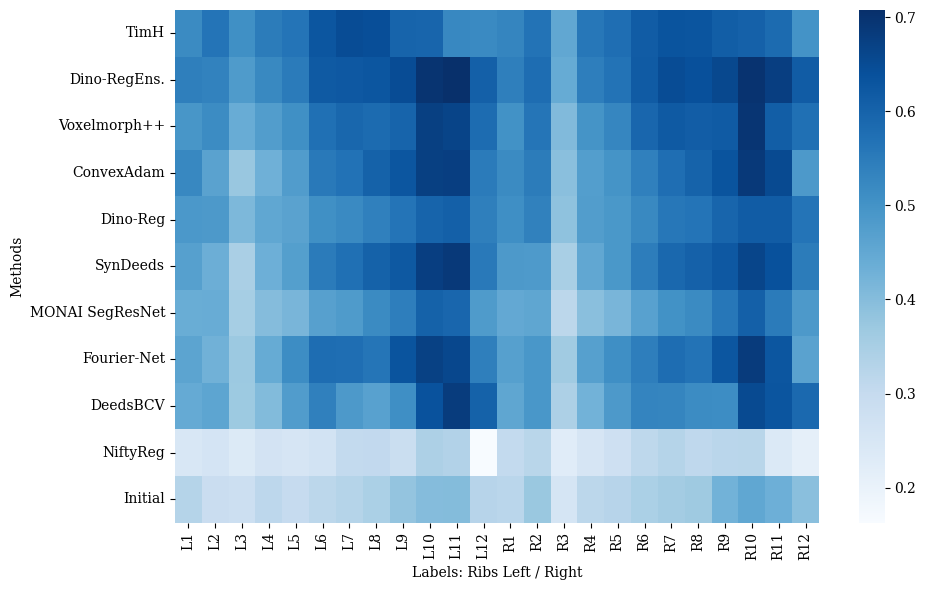}
		\caption{OncoReg: Dice \added{c}oefficients for individual rib-labels and each registration method, averaged across all \deleted{test} images.}
        \label{fig:oncoreg_ribs}
	\end{figure*}

\end{document}